\documentclass[american,aps,pra,reprint,floatfix,nofootinbib,superscriptaddress]{revtex4-1}
\usepackage[unicode=true,pdfusetitle, bookmarks=true,bookmarksnumbered=false,bookmarksopen=false, breaklinks=false,pdfborder={0 0 0},backref=false,colorlinks=false]{hyperref}
\hypersetup{colorlinks,linkcolor=blue,citecolor=red,urlcolor=cyan}
\usepackage{graphics,epstopdf,amsthm,amsmath,amssymb,mathptmx,colortbl,color,bm,framed,mathrsfs}
\usepackage{comment}
\usepackage{csquotes}
\usepackage[T1]{fontenc}
\usepackage[up]{subfigure}
\usepackage[braket, qm]{qcircuit}
\usepackage{physics}
\usepackage{commath}
\usepackage{mathtools}
\usepackage{float}
\usepackage{parskip}
\usepackage[demo]{graphicx}
\usepackage{centernot}
\usepackage[normalem]{ulem}
\usepackage{csquotes}

\begin{document}
%\emergencystretch 3em
\title{Inter-generational comparison of quantum annealers in solving hard scheduling problems}

\author{Bibek Pokharel}
\email{pokharel@usc.edu}
\affiliation{Department of Physics and Astronomy, and Center for Quantum Information Science \& Technology, University of Southern California, Los Angeles, California 90089, USA}

\author{Zoe Gonzalez Izquierdo}
\affiliation{QuAIL, NASA Ames Research Center, Moffett Field, California 94035, USA}
\affiliation{USRA Research Institute for Advanced Computer Science, Mountain View, California 94043, USA}

\author{P. Aaron Lott}
\affiliation{QuAIL, NASA Ames Research Center, Moffett Field, California 94035, USA}
\affiliation{USRA Research Institute for Advanced Computer Science, Mountain View, California 94043, USA}

\author{Elena Strbac}
\affiliation{Standard Chartered Bank, London, EC2V 5DD, UK}

\author{Krzysztof Osiewalski}
\affiliation{Standard Chartered Bank, London, EC2V 5DD, UK}

\author{Emmanuel Papathanasiou}
\affiliation{Standard Chartered Bank, London, EC2V 5DD, UK}

\author{Alexei Kondratyev}
\affiliation{Abu Dhabi Investment Authority (ADIA), P.O. Box 3600, Abu Dhabi, UAE}
\affiliation{Department of Mathematics, Imperial College London,  London, SW7 2AZ, UK}

\author{Davide Venturelli}
\affiliation{QuAIL, NASA Ames Research Center, Moffett Field, California 94035, USA}
\affiliation{USRA Research Institute for Advanced Computer Science, Mountain View, California 94043, USA}

\author{Eleanor Rieffel}
\affiliation{QuAIL, NASA Ames Research Center, Moffett Field, California 94035, USA}

\date{\today}

\begin{abstract}

We compare the performance of four quantum annealers, the D-Wave \emph{Two}, \emph{2X}, \emph{2000Q}, and \emph{Advantage} in solving an identical ensemble of a parametrized family of scheduling problems. 
These problems are NP-complete and, in fact, equivalent to vertex coloring problems. They are also practically motivated and closely connected to planning problems from artificial intelligence. We examine factors contributing to the performance differences while separating the contributions from hardware upgrades, support for shorter anneal times, and possible optimization of ferromagnetic couplings. While shorter anneal times can improve the time to solution (TTS) at any given problem size, the scaling of TTS with respect to the problem size worsens for shorter anneal times. In contrast, optimizing the ferromagnetic coupling improves both the absolute TTS and the scaling. There is a statistically significant improvement in performance between D-Wave Two and 2X and from all older generation annealers to Advantage, even when operated under identical anneal time and ferromagnetic couplings. However, the performance improvement from 2X to 2000Q requires the anneal time and ferromagnetic couplings to be optimized. Overall, owing to these inter-generational hardware improvements and optimizations, the scaling exponent reduces from $1.01 \pm  0.01$ on Two to $0.173 \pm 0.009$ on Advantage. 

\end{abstract}
\maketitle

\section{Introduction}
\label{sec:intro}

Theoretically, quantum computers are superior to classical computers when quantum mechanical laws are carefully exploited in deterministic algorithms~\cite{RPbook,NCbook}. However, in real-world deployment, empirical performance can take precedence over theoretical superiority. In fact, for most classical algorithms that perform well empirically on practical tasks, no proof exists for estimates or bounds on their performance.  For this reason, empirical experimentation of quantum heuristics is critical to understanding the breadth of quantum computing's impact. 

Quantum annealing is one of the most prominent quantum metaheuristic for optimization~\cite{Rieffel14CaseStudy,Venturelli15,kim2019,Marshall19_Pausing,gonzalez2021,Farhi98,Smelyanskiy12}. Quantum annealing has become accessible to experimentation in recent years, and the company D-Wave has produced and commercialized multiple devices in the last decade. Starting with the 128-qubit D-wave One, each generation of annealer has aimed to improve performance on optimization problems. The D-Wave Two, 2X, 2000Q, and Advantage systems (the four we use in this study) have 512, 1152, 2048, and 5640 qubits, respectively. The D-Wave 2000Q introduced several features like a shorter minimum anneal time, increased control of the annealing schedule, and a wider range of ferromagnetic coupling strengths. Advantage, the newest device, uses a denser native graph structure with 15 instead of 6 degrees of coupling between qubits. In addition to these increased number of qubits and features, these newer generations of annealers also have reduced noise and better calibration and reduction in analog control errors~\cite{advantage,dwave_ice}. \\

This report evaluates how the various updates made to these machines over the last few years impact their effectiveness in solving hard problems of practical interest. While the comparative performance of Advantage and 2000Q was discussed in~\cite{Willsch21} for exact cover problems, to our knowledge, this is the first investigation of its kind that compares the performance of four different generations of quantum annealers with respect to an identical set of applied problems. Our results can be summarized as follows:
\begin{itemize}
\item  Under the default operating conditions and fixed anneal times, 2X outperformed Two, and Advantage outperformed all its predecessors. While 2000Q was able to solve larger problems than 2X, there was no statistical difference in the average TTS between the two machines. Improvement from Two to 2X is likely due to reduced specification errors and other hardware level improvements. For Advantage, a large part of the performance increase can be attributed to increased connectivity. 
\item  The shortest possible anneal time gives the shortest TTS for most problem sizes, with the effect being most pronounced for small-sized problems. However, a longer anneal time is necessary to obtain solutions for the largest problem sizes that are addressable only via the Advantage platform. Crucially, the scaling of the TTS with respect to size increases as we decrease the anneal time. Overall, in future architectures with lower anneal times, problems of different sizes might benefit from size-appropriate anneal times.
\item In general, optimizing the ferromagnetic coupling beyond the default settings lowers the TTS at each problem size. However, the magnitude of this improvement is device-dependent. On 2X and 2000Q, the change in TTS due to such optimization is nearly negligible. However, on Advantage, there is a substantial reduction in the scaling exponent after ferromagnetic coupling optimization. While our brute-force optimization is informative, we suspect that as the devices and the problems increase in size, a more sophisticated method of optimizing the coupling will have to be employed.
\end{itemize}

The structure of the paper is as follows. Section \ref{sec:qaReview} briefly reviews quantum annealing.
Section \ref{sec:scheduling} describes the scheduling problem and the generation of problem instances. 
Section \ref{sec:methods} discusses the methodological details of our experiments. 
In section \ref{sec:results} we discuss our results, and summarize their implications in section \ref{sec:concl}. 
 
\begin{figure*}[t!]
    \centering
	\includegraphics[trim = 200 30 250 10, clip, scale=0.5]{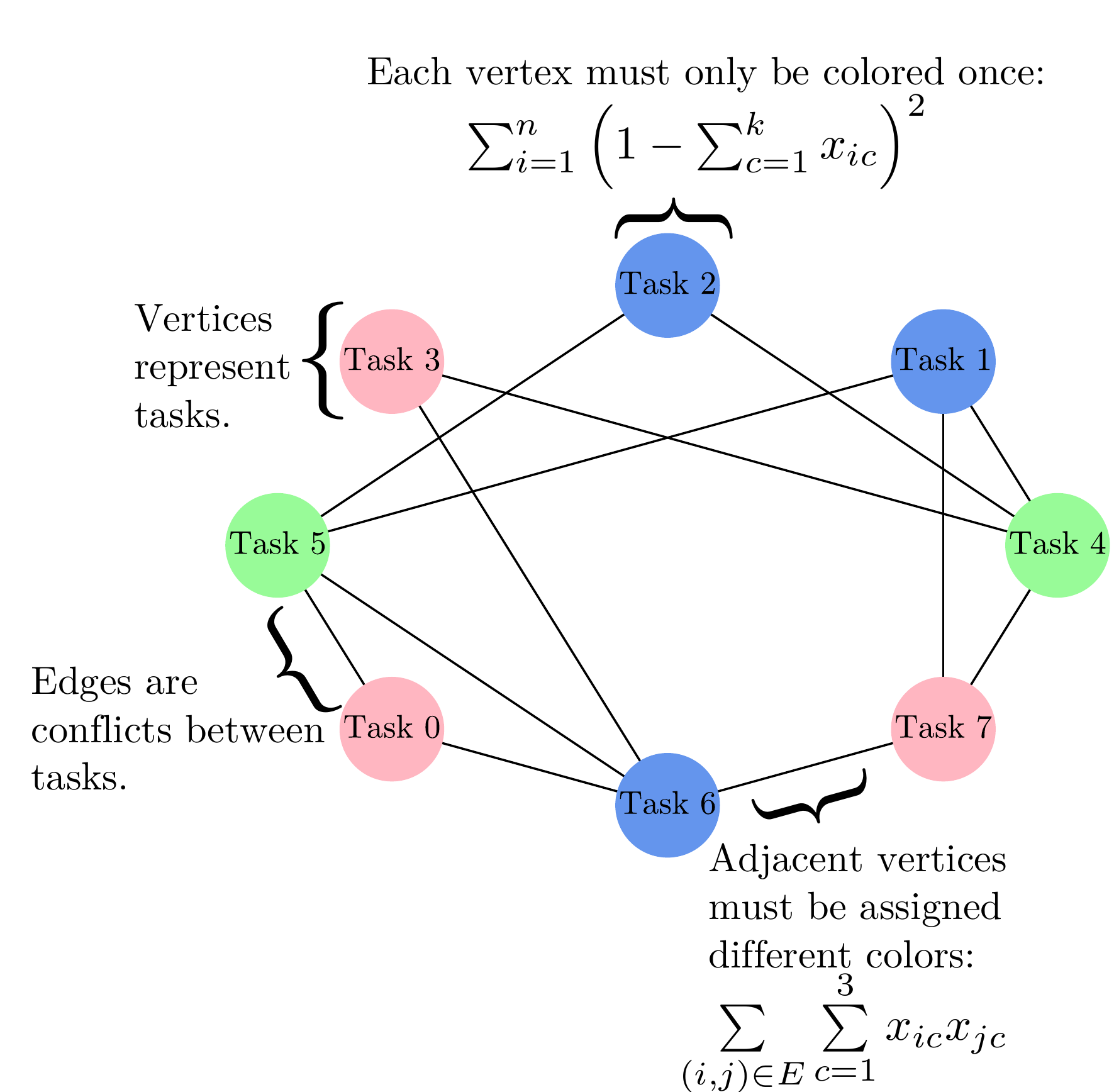}
	\caption{Example of a schedule for n=8 is shown with a valid 3-coloring. Each scheduling problem can be mapped to a graph-coloring problem by identifying tasks with nodes and task conflicts as edges. }
\label{fig:coloring_example}
\end{figure*}

\section{Review of quantum annealing}
\label{sec:qaReview}

In quantum annealing, the device starts by implementing an initial Hamiltonian $H_I$, and the overall Hamiltonian $H$ is parametrically changed until it becomes the final Hamiltonian $H_F$. This process corresponds to the following Hamiltonian evolution
\begin{equation}
H(s) = A(s) H_I + B(s) H_F \ , \ s \in [0,1], 
\label{eq:annealing}
\end{equation} 
where $A(s)$ is a monotonically decreasing function and $B(s)$ is a monotonically increasing function such that $A(1)=B(0)=0$. $H_F$ encodes the cost-function landscape of a combinatorial optimization problem in its eigenvalues and its ground state maps to the optimal solutions to the problem. Quantum annealing utilizes both thermal and quantum fluctuations when exploring the cost-function landscape. So, in principle, it is more expressive and possibly more effective than simulated annealing or parallel tempering methods, where quantum effects like tunneling are not possible~\cite{finnila1994quantum}.  Due to practical aspects such as finite anneal time, temperature, noise, and sub-optimal parameter setting, the final output of an annealing run might not necessarily be a low-energy state of the problem Hamiltonian. Consequently, quantum annealers are usually treated as black-box optimizers that need to be iteratively run in conjunction with a parameter setting strategy.  In benchmarking studies, the data collected from multiple anneals are used to calculate the probability of the ground state solution
\begin{align}
P_{\text{gs}} = \frac{\text{Number of ground state solutions}}{\text{Total anneals}}.
\end{align}
Its counterpart, the time-to-solution (TTS) is defined as the expected time to obtain the ground state solution with 0.99 success probability for a specific anneal time, and is computed as 
\begin{align}
\text{TTS} = \frac{\ln(1-0.99)}{\ln(1-P_{\text{gs}})} t,
\end{align}
where $t$ is the anneal time for a single annealing repetition. 

The D-Wave quantum annealers have $2$-local architectures, meaning that there are pairwise couplings between the qubits, i.e., the Hamiltonian $H_F$ is an Ising Hamiltonian. Consequently, the optimization problem is formulated as a  quadratic unconstrained binary optimization (QUBO) problem \cite{Choi08,Lucas13,Shin2014comment},  where the cost function is a quadratic function of binary variables. We will refer to obtaining the QUBO for a problem as \emph{mapping} the problem to its QUBO.  As each qubit in the quantum annealer is connected to a small subset of the other qubits, representing a single binary variable usually requires multiple physical qubits to reproduce the QUBO's connectivity faithfully. This set of physical qubits representing a single logical variable is called a vertex model, and the process of finding vertex models for each logical qubit in the problem is called \emph{embedding} of the QUBO in the hardware chip. The physical qubits within a vertex model are subjected to ferromagnetic coupling $J_F$ in the final Hamiltonian.  Finding the value $J_F$ that maximizes the TTS is called the \emph{parameter-setting} problem~\cite{Pudenz2016}. For a more detailed review of quantum annealing on D-Wave machines, the reader is invited to read Ref.~\cite{Johnson2011quantum,job2018test,Boothby2021}. 

\section{Parameterized families of scheduling problems}
\label{sec:scheduling}

Scheduling problems deal with the allocation of time and resources under certain constraints.  While these NP-complete problems are important on their own, they also have many applications to planning problems from artificial intelligence \cite{chien:asp}. Here, we consider the task of assigning $k$ time-slots to $n$ tasks while avoiding any possible time conflicts (double-scheduling). These problems are equivalent to vertex-coloring problems, which are also a class of NP-complete problems. In particular, when we map the time-slots to colors and tasks to vertices, the edges between the vertices correspond to constraints between the tasks (see Fig. \ref{fig:coloring_example}). So the colorability of the graph refers to the solvability of the scheduling problem. The $k$-coloring gives the desired conflict-free schedule.

The coloring task can be represented using doubly-indexed binary variables $x_{ic}$, where $x_{ic}=1(0)$ means that the $i$th vertex is colored (not colored) with color $c$. Each vertex must be colored only once, and adjacent vertices must have different colors. Using these two conditions, we get the QUBO:
\begin{equation}
	H(\vec{x})=\sum\limits_{i=1}^{n} \left( 1- \sum\limits_{c=1}^{k} x_{ic} \right)^2 + \sum\limits_{(i,j)\in E} \sum\limits_{c} x_{ic} x_{jc}	
    \label{eq:graphColoringQUBO}
\end{equation}
For a valid coloring, this objective function will give $H(\vec{x})=0$.

We consider an ensemble of problems that are not easily colorable or trivially uncolorable. To be more specific, we consider the 3-colorability of a specific set of  Erdos-Renyi graphs $G_{n,p}$.  $G_{n,p}$ are graphs with $n$ vertices such that the probability of having an edge between each pair of vertices is $p$. A combinatorial phase transition for $k$-colorability problems is known in terms of the parameter $d = \frac{e}{n}$ where $e$ is the number of edges \cite{Achlioptas99}. This phase transition occurs when the difficulty of finding the coloring goes from easy to hard to easy. This easy-hard-easy transition of colorability is well studied, and there are upper and lower bounds on the transition parameter $d$. However, finding the exact location of the phase transition is still an open problem \cite{CojaOghlan13}. We use $d = 4.5$ for the generation of our graphs, following the study in~\cite{Rieffel2014AAAI}. We generate these graphs using a C++ program, which is an extension of the graph-generation method used by Culberson et al. \cite{Culberson95hidingour}. The graphs used in this paper are identical to the set used by Rieffel et al. \cite{casestudy} in their case study about the performance of quantum annealers in solving planning-type problems.

\section{Methods}
\label{sec:methods}

\begin{table*}
\centering
 \begin{tabular}{||c c c c c ||} 
 \hline
 D-Wave &  Two & 2X & 2000Q & Advantage \\ [0.5ex] 
 \hline\hline
QPU Image &  \includegraphics[height=1.8cm]{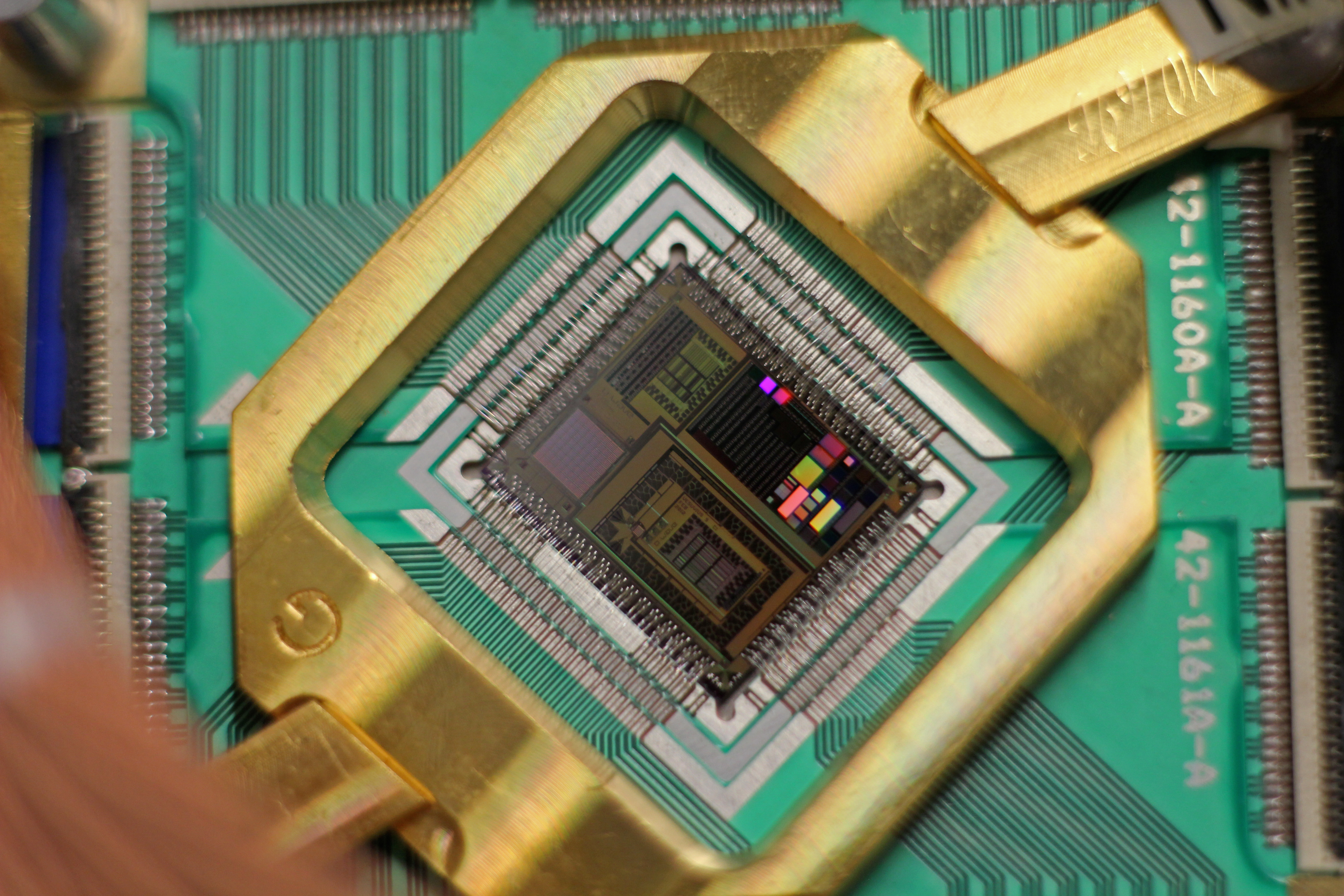} & \includegraphics[height=1.8cm]{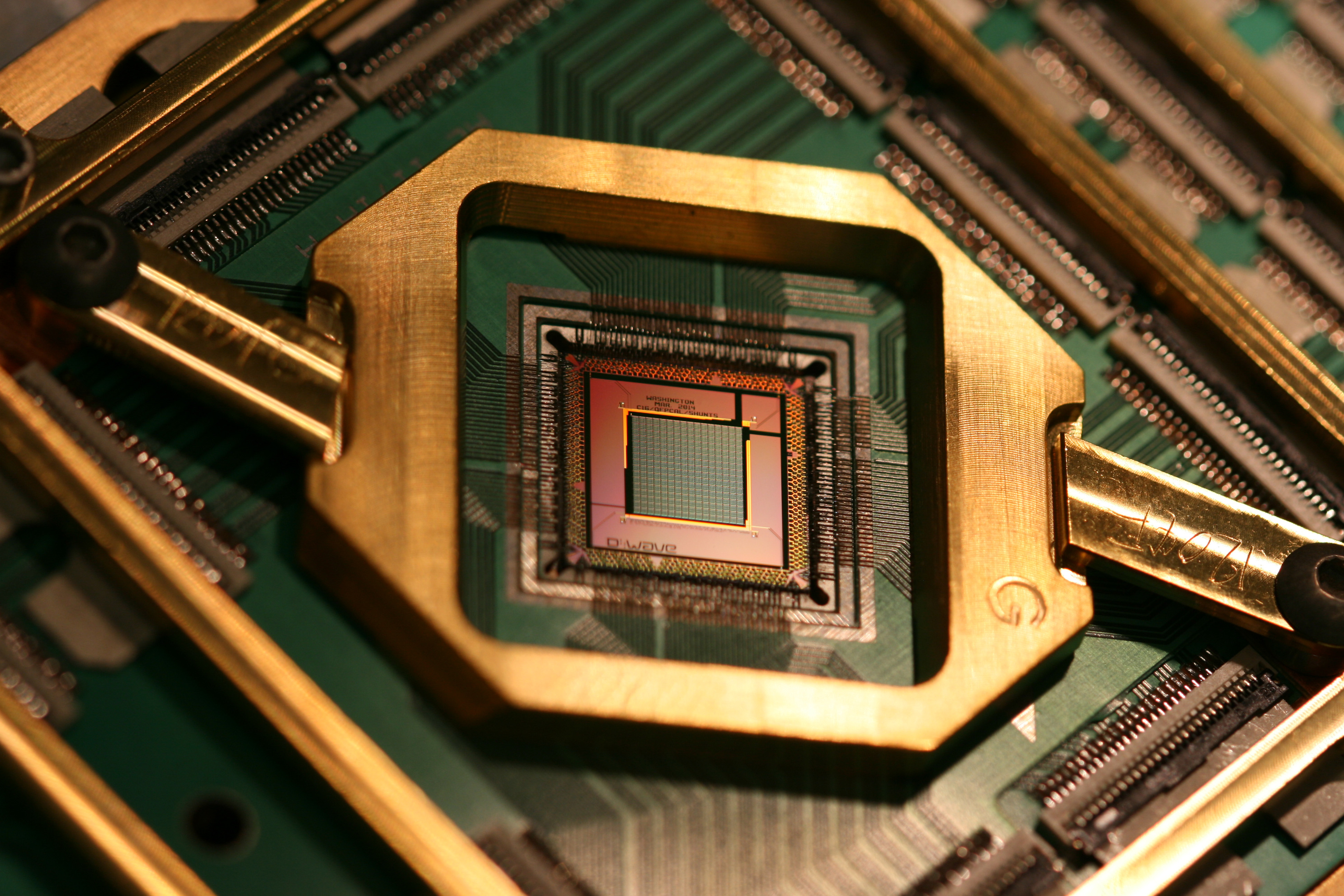} & 	\includegraphics[height=1.8cm]{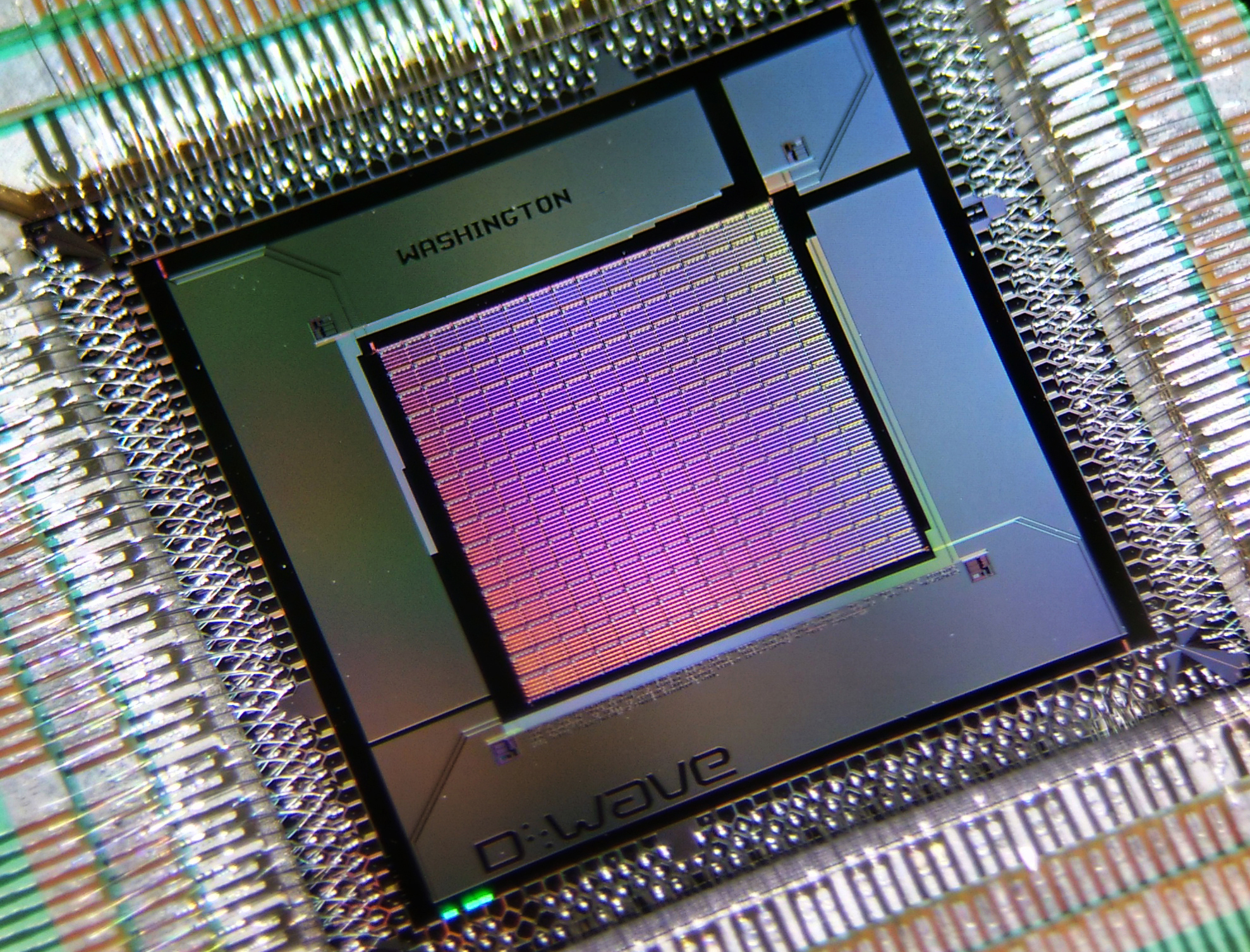} & \includegraphics[height=1.8cm]{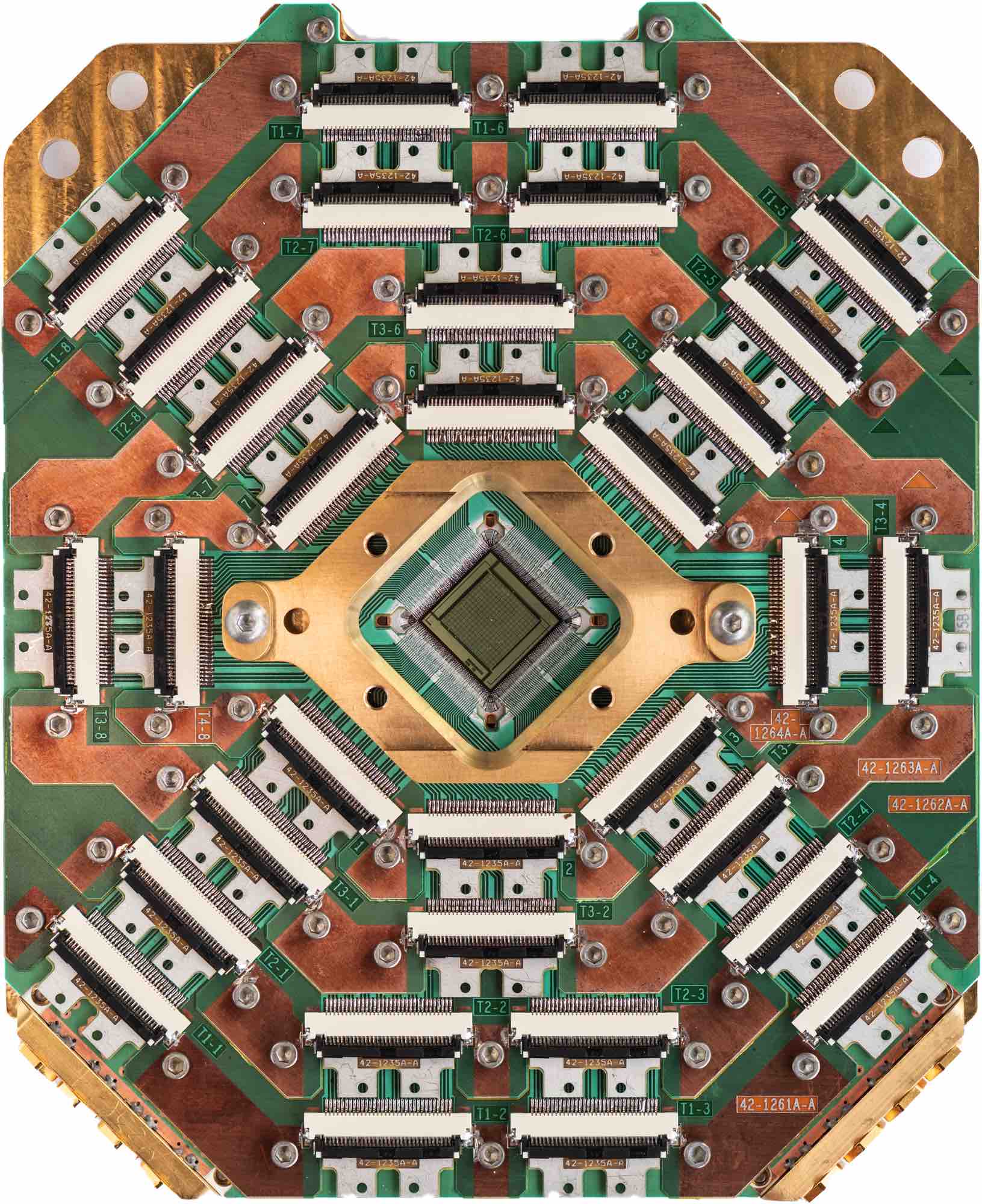} \\ [0.5ex] 
 \hline
 Total Qubits & 512 & 1152 & 2048 & 5640 \\
 \hline
 Working Qubits & 509 & 1097 & 2031 & 5436 \\
 \hline
 Architecture & Chimera & Chimera & Chimera & Pegasus \\
 \hline 
  Min Anneal time & 20 $\mu$s & 5 $\mu$s & 1 $\mu$s & 1 $\mu$s \\
 \hline 
Temperature & 13.2 mK & 12 mK & 12.1 mK & 15.8 $\pm$ 0.5 mK \\
 \hline
 Anneals per gauge & 45000 & 10000 & 10000 & 500 \\
  \hline
 Problem Size & 8-16 & 8-20 &  $\begin{array}{cc} \text{8-24} & \text{not optimized} \\ \text{8-20} & \text{optimized} \end{array}$ & 8-40 \\
 \hline
 Gauges & 10 & 10 &  $\begin{array}{cc} 100 & n=20,22,24 \\ 50 & n=18 \\ 10 & \text{otherwise} \end{array}$ & 20-400 depending on $J_F$, $n$ and $t$ \\

\hline

 \end{tabular}

\caption{Hardware details for all four generations of the D-Wave annealers, the first three hosted at NASA Ames Research Center and the last one accessed through the D-Wave Leap cloud platform, are shown. The images of the chips were provided by D-Wave Systems. Note that the number of working qubits for a specific device is typically slightly lower than the number of total qubits since fabrication defects leave a small portion of components unusable. This table also lists the default problem sizes, gauges, number of anneals, and anneal time used for each D-Wave annealer.}
\label{tb:hardware}

\end{table*}

\begin{figure*}[t!]
\centering
	\includegraphics[trim = 20 20 20 20, clip,width=0.42\linewidth]{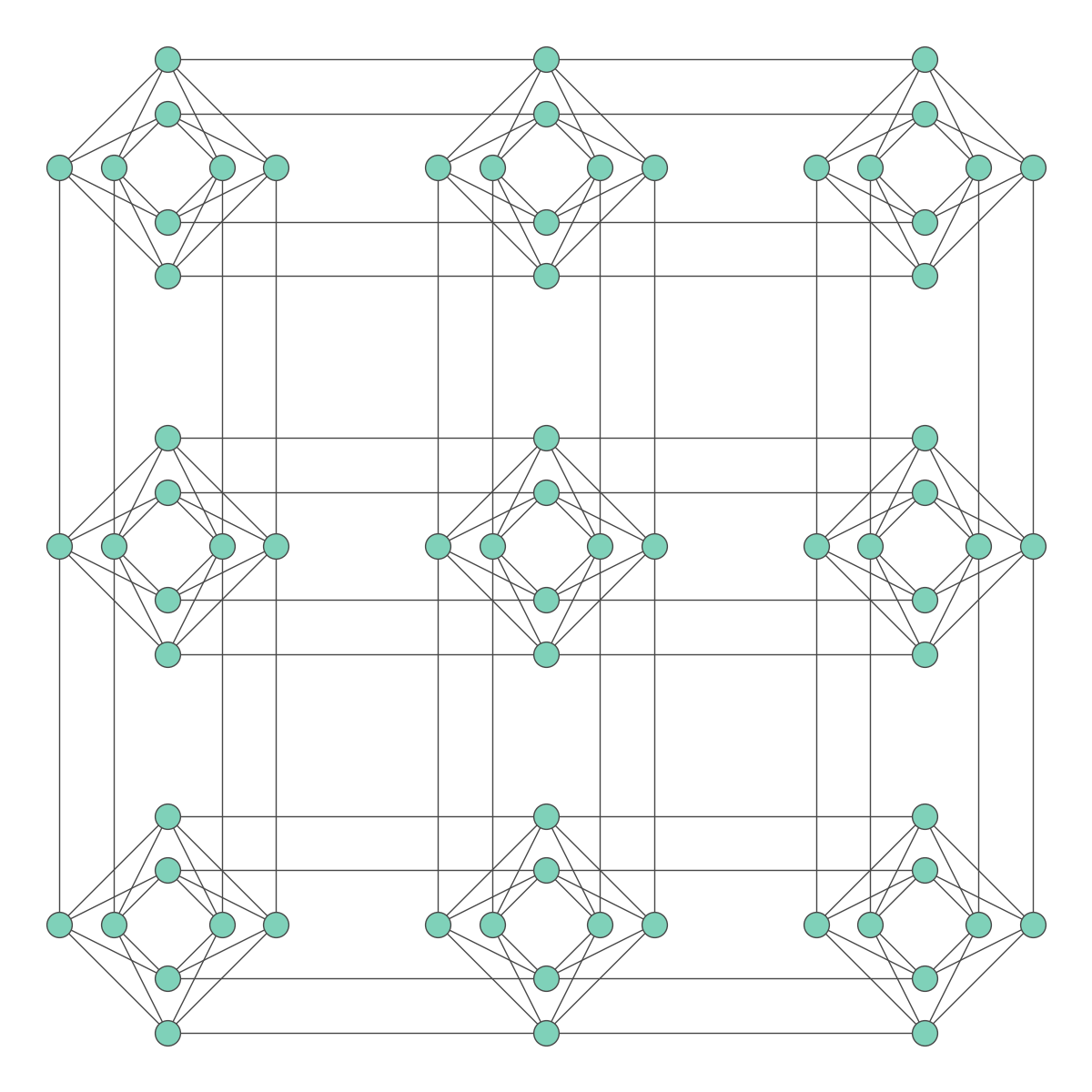}
	\includegraphics[trim = 20 20 20 100, clip,width=0.57\linewidth]{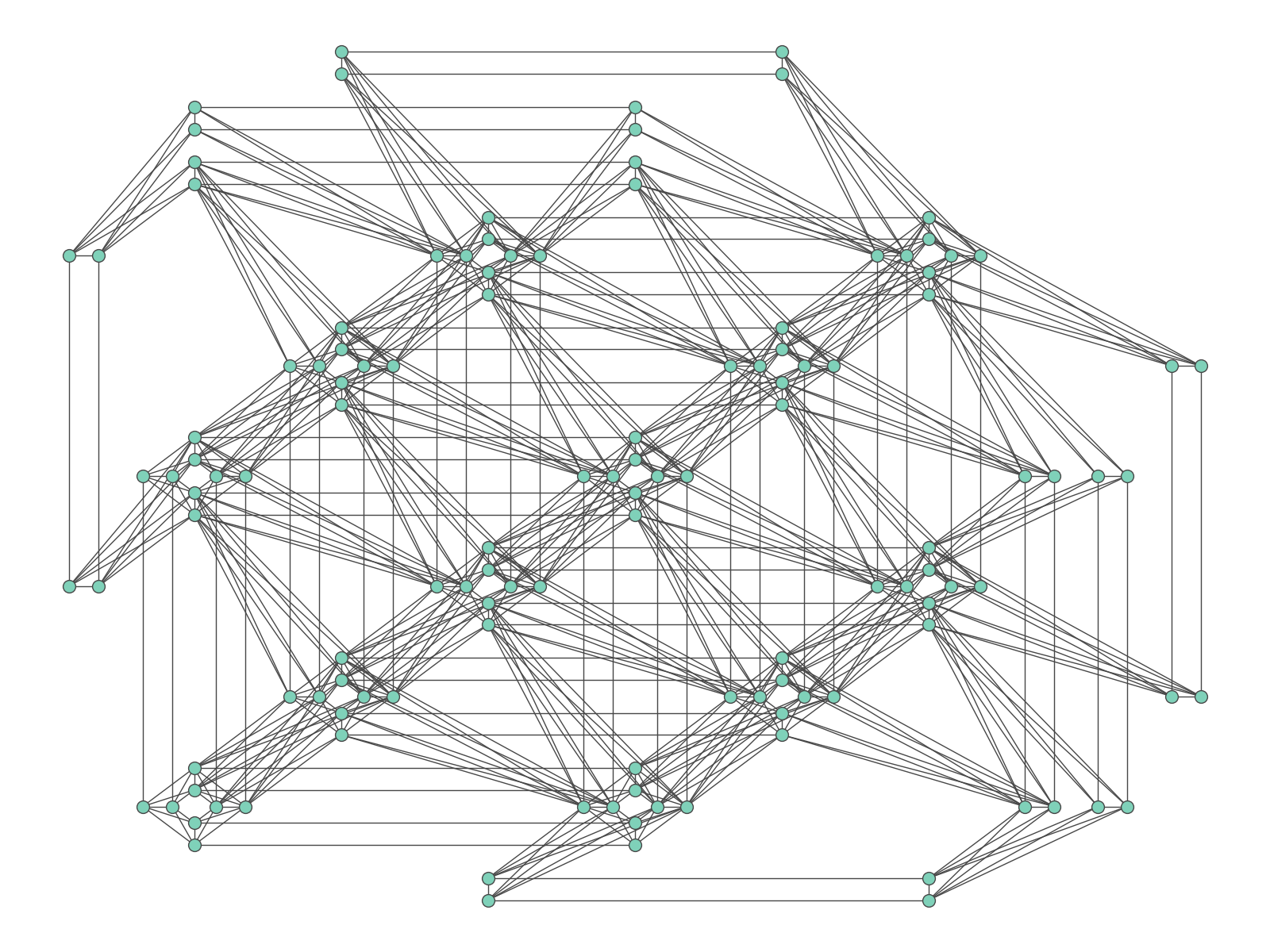}
	\caption{\textbf{Left:} Structure of the Chimera graph used by D-Wave Two, 2X, and 2000Q. 8-qubit, bipartite cells are arranged in a square pattern. Shown is a C3 graph; with three cells per side, for a total of 9 cells and 72 qubits. The pattern repeats to attain larger numbers of qubits: C8 for Two, C12 for 2X, and C16 for 2000Q. \textbf{Right:} Pegasus graph used by D-Wave Advantage. Shown is a P3 graph, with 12 full cells and several partial ones. Like with Chimera, the pattern repeats to create the larger P16 graph that Advantage features. Notice that Chimera is a subgraph of Pegasus; any problem native to Chimera is also native to Pegasus.}
	\label{fig:pegasus_chimera}
\end{figure*}

Quantum annealing runs were performed on the three generations of D-Wave quantum annealers (Two, 2X, and 2000Q) housed at NASA Ames Research Center and the latest generation (Advantage), accessed through the D-Wave Leap cloud platform. The number of qubits, minimum anneal times, and base operating temperature for all four annealers are listed in Table. \ref{tb:hardware}.  We used 100 problem instances at each problem size while restricting the instances to ones with at least one valid schedule (i.e., coloring). We found the embeddings for the QUBO instances using D-Wave's native heuristic \verb|find_embedding|~\cite{Cai-14}. 

To mitigate biases toward certain solutions formed by asymmetries in the processors, we used random gauges, which specify whether {0,1} bits are mapped to {-1,1} or {1,-1}. Table. \ref{tb:hardware} outlines the number of gauges used for different problem sizes. We check for the number of valid schedules (i.e., colorings) obtained by using Eq. \ref{eq:graphColoringQUBO} and report the corresponding median TTS.  We bootstrap the median TTS from 100 problem instances to get 5000 samples. The reported TTS is the mean of this ensemble, and the error bars correspond to 95\% confidence intervals. Scheduling problems are NP-complete problems, so we expect our TTS to scale exponentially with size in the asymptotic limit. This exponential scaling is expected both for quantum and classical algorithms, with the scaling varying between different algorithms \cite{Rieffel2014AAAI}.  So we fit $\text{TTS}= T_0 e^{n \alpha}$~\cite{Kowalsky21} and report the scaling exponent $\alpha$.

Each generation of D-Wave annealers has a larger chip layout structure than their predecessor, increasing the size of problems that can be embedded in each device. Moreover, the Advantage platform has transitioned the topology from a \emph{Chimera} graph to a \emph{Pegasus} graph~\cite{pegasus_topology,advantage} (see Fig.~\ref{fig:pegasus_chimera}), which has a higher connectivity. We were able to embed problems with up to 40 tasks on D-Wave Advantage and D-Wave 2000Q, and up to 32 tasks on D-Wave 2X. However, problems of sizes larger than 16, 20, and 24 on Two, 2X and 2000Q respectively had $P_{gs} = 0$. For each machine, the last plotted dot in the curve indicates the largest problem size for which the statistics analysis returned a non-infinite median. Working with larger problem sizes would require millions of runs per instance to obtain sufficient statistics. 
 
In line with the typical practice for applied problems~\cite{Pudenz2016,gonzalez2021}, we optimized the ferromagnetic couplings. For this optimization, we set the anneal time to 5, 1, and 20 $\mu$s for 2X, 2000Q, and Advantage, respectively. In particular, for each machine, we chose the anneal time that gave the shortest TTS for the largest problem size we considered before $J_F$ optimization. When optimizing the ferromagnetic coupling $J_F$ for 2X and 2000Q, we chose $J_F \in [-0.5, -2.0]$ and $[-0.625, -1.375] $ with steps of 0.125, respectively.  For both machines, relatively few problem sizes had their optimal coupling at  $|J_{F}|>1.0$. $J_{F}=-2.0, -1.375$ were not the optimal coupling for any problem sizes on 2X and 2000Q, respectively.  In our plots, `i.opt' refers to $J_F$ optimized for each problem instance separately. A similar optimization of $J_F$ was done for D-Wave Two by Rieffel et al. in \cite{casestudy}, so we do not repeat those experiments here. For 2000Q data, we restrict $J_F$ optimization to $n<20$ due to limited computational resources. Moreover, for 2X and 2000Q, we performed coupling optimization at various anneal times and found that the optimal coupling did not depend on the anneal time. Optimization of $J_F$ on Advantage has been studied in detail in other work currently in preparation by some of the same authors. Here we explored the range $J_F \in [-0.4, -1]$ for $n=24$ and $t=20 \mu$s and found $J_F = -0.5$ to be optimal (labeled `opt'). Unless specified otherwise, the default problem sizes, gauges, number of anneals, and anneal times used are detailed in Table~\ref{tb:hardware}.  We discuss the results in the next section.

\section{Results}
\label{sec:results}
\label{sec:default}

\begin{figure}[th!]
	\includegraphics[trim = 5 0 0 0, clip, width=\columnwidth]{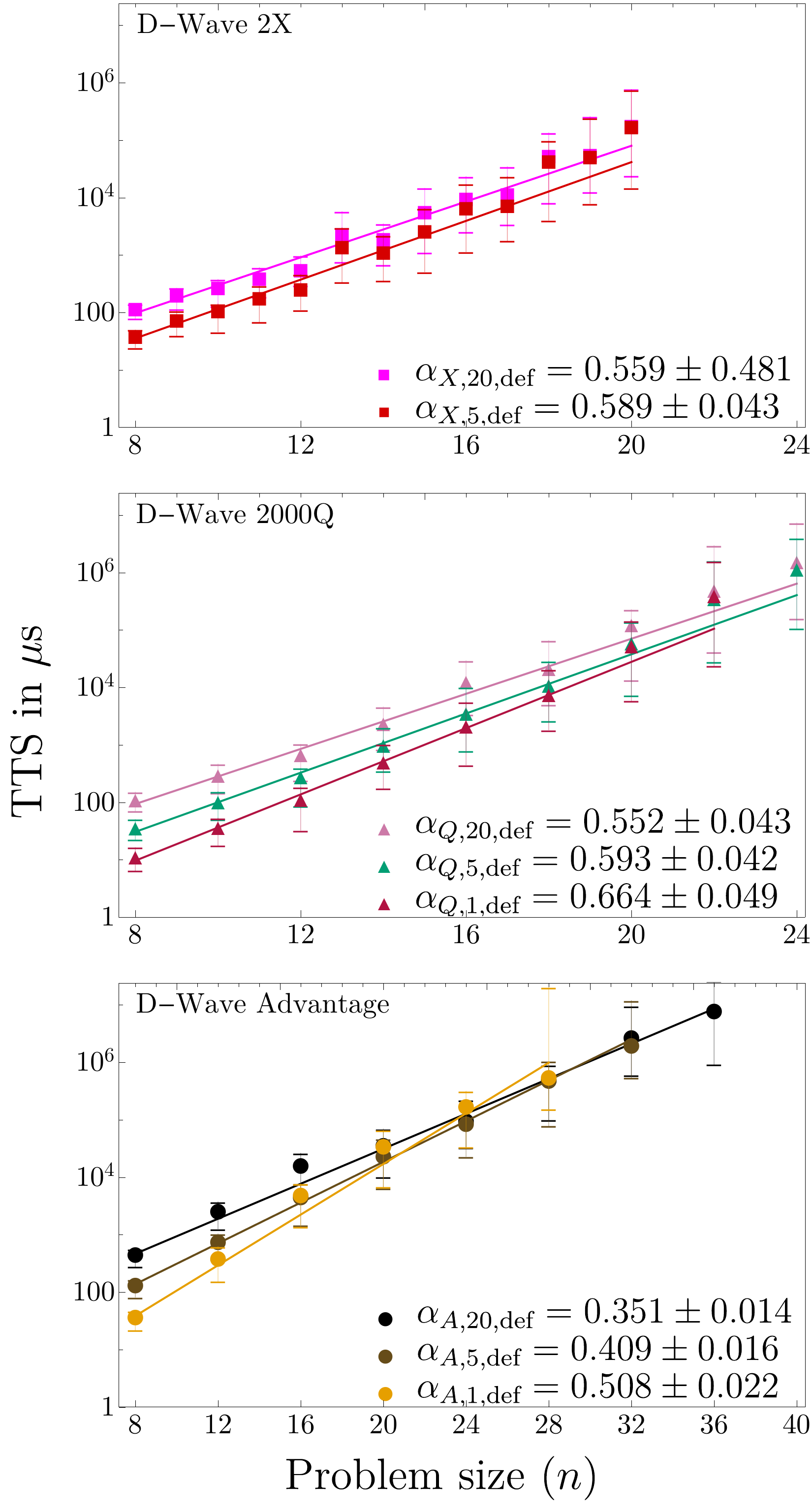}
	\caption{TTS as a function of problem-size across generations at various anneal times. Ferromagnetic coupling is set to the default value. The scaling exponent is labeled as $\alpha_{\text{machine, anneal time, $J_F$ optimization}}$.}
\label{fig:ta_comparison}
\end{figure}

\begin{figure}[th]
	\includegraphics[trim = 6 0 0 0, clip, width=\columnwidth]{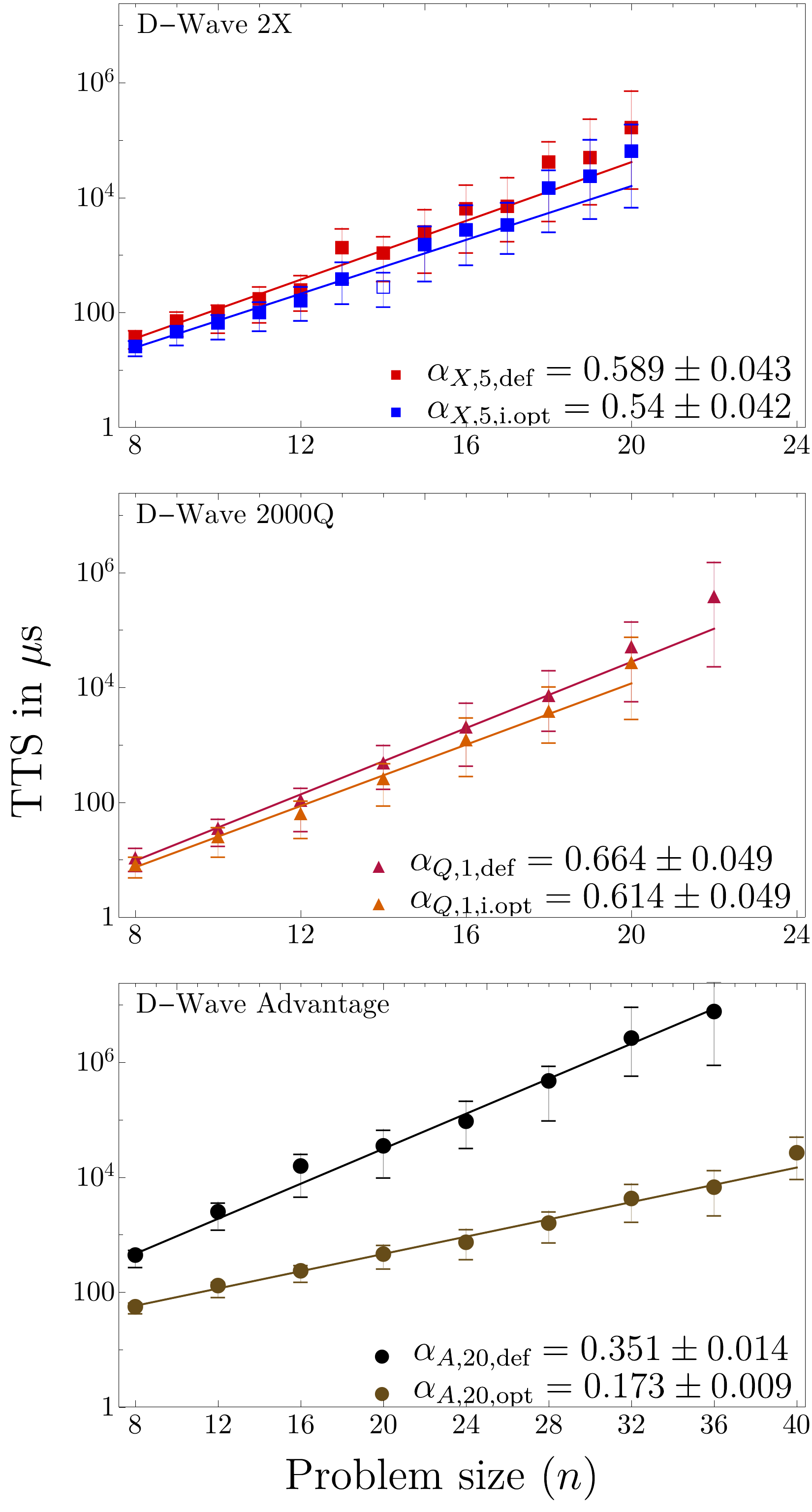}
	\caption{TTS as a function of problem-size across generations. $J_F$ is optimized at the best performing anneal time. Optimizing $J_F$ both reduces the TTS at each size and the exponent $\alpha$. The scaling exponent is labeled as $\alpha_{\text{machine, anneal time, $J_F$ optimization}}$.}
\label{fig:jf_comparison}
\end{figure}

\begin{figure*}[th]
	\includegraphics[trim =0 0 0 0, clip, width=\linewidth]{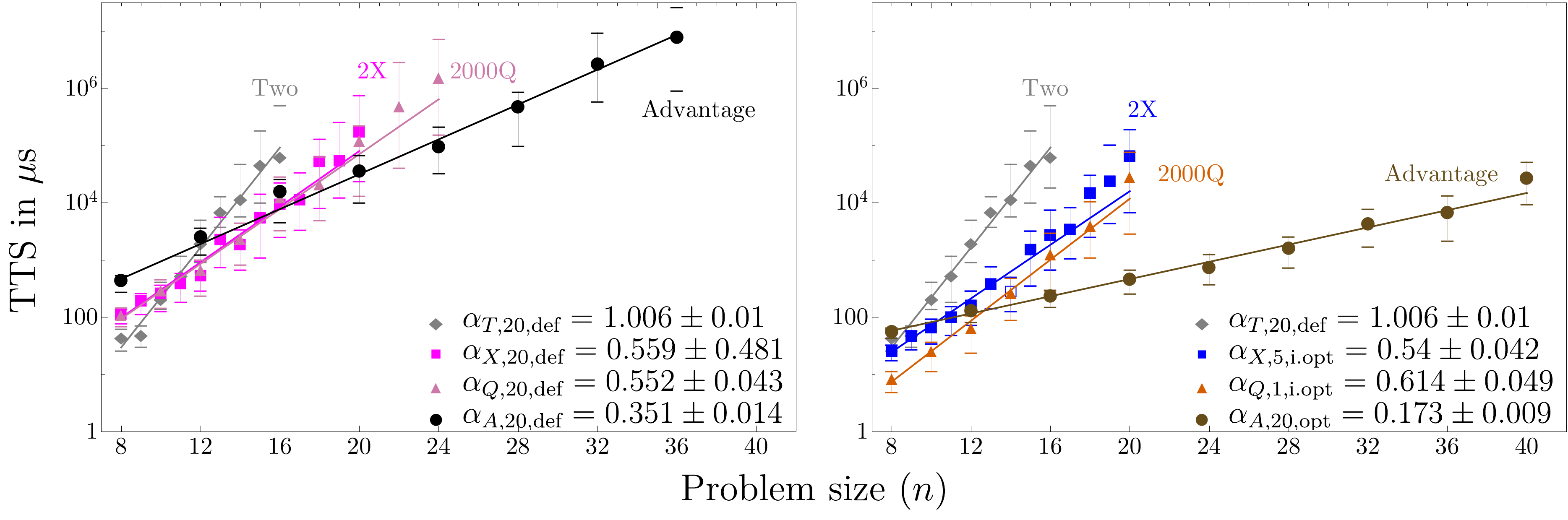}
   \caption{Left: TTS as a function of problem-size across generations. The time and $J_F$ are set to the default value. Right: anneal time, and $J_F$ are optimized for each machine. The scaling exponent is labeled as $\alpha_{\text{machine, anneal time, $J_F$ optimization}}$.
    } 
    \label{fig:mainresults}%
\end{figure*}

D-Wave's quantum annealers allow the user to vary the anneal time while keeping the weight functions in Eq.~\ref{eq:annealing} fixed. Using the shortest possible anneal time increases the number of solutions obtained in a fixed period, but it also affects the quality of the solutions \cite{casestudy}. We evaluate this tradeoff by considering TTS as a function of the anneal time and problem size.  Fig.~\ref{fig:ta_comparison} shows TTS as a function of problem size at various anneal times for D-Wave 2X, 2000Q, and Advantage. Generally, when varying $t$ while fixing the problem size $n$,  the lowest possible anneal time gives the lowest TTS for most graph sizes. However, the improvement is less significant for larger problem sizes, and the error bars on the TTS also increase with $n$. In other words, the improvement offered due to shorter anneal time becomes negligible for larger problem sizes. The scaling exponent $\alpha$ increases slightly for shorter anneal times. This effect is most pronounced in Advantage where the exponent goes from $\alpha_{\text{A,20$\mu$s,def}} = 0.351 \pm 0.014$ at 20 $\mu$s to $\alpha_{\text{A,1$\mu$s,def}} = 0.508 \pm 0.022$ at 1 $\mu$s. However, note that if we limit ourselves to $n \leq 22$ (like for 2000Q), Advantage's optimal annealing time would also be 1 $\mu$s, and it would outperform 2000Q for the smaller sizes. Overall, our results suggest that lowering anneal time further in future architectures could benefit small-sized scheduling problems. Future annealers with lower anneal times and larger problem sizes might benefit from assigning different anneal times to problems of different sizes. 

Generally, $J_F$, the ferromagnetic coupling applied to each vertex model, is set to be larger than the coupling between the logical qubits $J_{ij}$ and the local field $h_{i}$.  However, if $J_F$ is arbitrarily large, the coupled qubits may have the same spin, but the solution might not correspond to the ground state of the QUBO. By default, on D-Wave annealers, $J_F = -1.0$ (where $J_{ij}, h_i \in [-1.0,+1.0]$, with this range expanded for the latest generations). We considered optimizing $J_F$ across problem sizes and problem instances (except for Advantage, where we only optimized for $n=24$). 

In general, we found that setting  $J_F = -1.0$, as D-Wave's native algorithm does, is not the optimal setting.  Rieffel et al. \cite{casestudy} had argued for non-optimality of this default setting and further observed that for D-Wave Two, the ratio of the optimal ferromagnetic coupling $|J_F|$ to the maximum internal coupling $|J_{ij}|$ decreases with problem sizes for scheduling-type problems. We do not observed this trend on the newer machines. For 2X and 2000Q, $|J_{F}|=0.875$ and $0.75$ generally gives the shortest median TTS.  For Advantage, we found $-0.5$ to be the optimal value after exploring the range $J_F \in [-0.4, -1]$ for $n=24$ and $t=20 \mu$s.

As shown in Fig.~\ref{fig:jf_comparison}, the optimization of $J_F$ leads to a slight, statistically insignificant reduction in TTS for 2X and 2000Q.  Optimizing $J_F$ further across each vertex-model using standard optimization methods, like Nelder-Mead and gradient descent, was unwieldy. Due to the noisiness in the solution probability landscape, we observed that these search methods could not improve upon instance-level optimization. We suspect that as quantum annealers become less noisy, similar or more sophisticated optimization methods might succeed.

The effect of $J_F$ optimization is most pronounced for Advantage. Here for all sizes and annealing times, $J_F=-0.5$ performs better than the default setting. For example, the median TTS for the $n=40$ instances is infinity for all  annealing times (1,5, and 20 $\mu$s) with $|J_F|$ = 1, while it is finite for $|J_F| = 0.5$, and as low as $2.8 \times 10^{4}$ $\mu$s when the annealing time is 20 $\mu$s. Moreover, we also see the scaling exponent decrease by a factor of two before and after $J_F$ optimization. We suspect that this change has to do with the improved connectivity of the Pegasus architecture. As the vertex models are much smaller in Advantage compared to the older annealers, a strong $|J_F|$ is no longer necessary to keep these vertex models from breaking. Overall, our observations make a strong case for the non-optimality of the default $J_F$ setting. 

Lastly, we compare the performance of each subsequent generation of annealers by solving identical problems at default and optimized settings (see Fig. \ref{fig:mainresults}). The results in both configurations confirm the intuition that as the machine got upgraded, the performance improved. Note that the effect of hardware improvements between 2X and 2000Q was minimal. While optimizing the anneal time and $J_F$ was shown to reduce the TTS slightly, these minor improvements add up. Using the optimized $J_{F}$ at the lowest possible anneal time of 1 $\mu$s, we do observe shorter TTS in 2000Q than 2X. Besides the obvious opportunity to test larger problem sizes, we observe a change in slope for the median TTS, which is very appreciable between generations and especially striking on Advantage. This performance improvement is likely primarily due to the smaller vertex model sizes required for embedding a densely connected Ising in the Advantage chip, owing to the increased connectivity of the Pegasus graph.

\section{Conclusion}
\label{sec:concl}

We analyzed and quantified the comparative performance of four generations of quantum annealers, D-Wave Two, 2X, 2000Q, and Advantage in solving a parameterized family of hard scheduling problems.  By solving an identical ensemble of problems on each machine, we highlight how the hardware updates, lower anneal times, and the size and the connectivity of the architectures affect the time-to-solution for these problems. Under the default settings, we found a noticeable improvement in performance between 2X and Two, but not when updating from 2X to 2000Q. Advantage outperforms all of its predecessors.  

Using the shortest possible anneal time gave the lowest TTS at most problem sizes, but this improvement got less pronounced as the problem size increased. For the largest problem sizes, using a longer anneal time yielded the best results. Optimizing the ferromagnetic coupling across each problem instance also reduced TTS, but the improvement was not pronounced for 2X and 2000Q. On Advantage, we found a substantial reduction in TTS when using an optimized ferromagnetic coupling, both in absolute terms and the scaling exponent. Overall, the exponent improved by nearly a factor of six, from $1.006 \pm 0.01$ on Two to $0.173 \pm 0.009$ on Advantage. 

Hardware upgrades and optimization of operational parameters like anneal times and ferromagnetic couplings are crucial for improving quantum annealing performance. Using sophisticating annealing schedules, as demonstrated in ~\cite{Marshall19_Pausing,gonzalez2021}, are other ways to improve hardware performance that we did not consider in this report. Quantifying the performance of scheduling problems using these advanced annealing schedules would be a natural extension of this work. As the problems used here can be generated systematically for larger problem sizes \cite{Rieffel2014AAAI}, these problems serve as a valuable and fair method of benchmarking future quantum annealers, both against other annealers and state-of-the-art classical heuristics. 

% \vfill

\section*{Acknowledgements}

We want to thank Jeremy Frank, Minh Do, and NASA QuAIL team members for their valuable input throughout this investigation. B.P. did a significant portion of this work as a graduate student at the University of New Mexico (UNM) and as an intern at NASA and Stinger Ghaffarian Technologies (SGT), so he wants to thank UNM Physics and Astronomy Department and SGT for their support.  Recent work of ZGI, DV, and PAL is supported by USRA NASA Academic Mission Service (NNA16BD14C). The NASA QuAIL team also appreciates the support from the AFRL Information Directorate under grant F4HBKC4162G001, DARPA under IAA 8839 annex 125, and the Office of the Director of National Intelligence (ODNI) and the Intelligence Advanced Research Projects Activity (IARPA), via IAA 145483. The views and conclusions contained herein are those of the authors and should not be interpreted as necessarily representing the official policies or endorsements, either expressed or implied, of ODNI, IARPA, AFRL, or the U.S. Government. The U.S. Government is authorized to reproduce and distribute reprints for Governmental purposes, notwithstanding any copyright annotation thereon. The computer time on D-Wave Leap for the Advantage system was provided thanks to the support of Standard Chartered Bank.

\bibliographystyle{spphys}    
\bibliography{master,phaseTrans,qc}  

{\catcode`\/=\active \catcode`\.=\active \catcode`\-=\active
  \catcode`\@=\active \gdef\url{\tt\catcode`\/=\active \catcode`\.=\active
  \catcode`\-=\active \catcode`\@=\active
  \def/{\discretionary{\char`\/}{}{\char`\/}}%
  \def.{\discretionary{\char`\.}{}{\char`\.}}%
  \def-{\discretionary{\char`\-}{}{\char`\-}}%
  \def@{\discretionary{\char`\@}{}{\char`\@}}}} \def\annote#1{} %{[#1]}
  \def\tilde{\char126}
\begin{thebibliography}{10}
\providecommand{\url}[1]{{#1}}
\providecommand{\urlprefix}{URL }
\expandafter\ifx\csname urlstyle\endcsname\relax
  \providecommand{\doi}[1]{DOI \discretionary{}{}{}#1}\else
  \providecommand{\doi}{DOI \discretionary{}{}{}\begingroup
  \urlstyle{rm}\Url}\fi

\bibitem{RPbook}
E.G. Rieffel, W.~Polak, \emph{A Gentle Introduction to Quantum Computing} (MIT
  Press, Cambridge, MA, 2011)

\bibitem{NCbook}
M.~Nielsen, I.L. Chuang, \emph{Quantum {C}omputing and {Q}uantum {I}nformation}
  (Cambridge University Press, Cambridge, 2001)

\bibitem{Rieffel14CaseStudy}
E.G. Rieffel, D.~Venturelli, B.~O'Gorman, M.B. Do, E.M. Prystay, V.N.
  Smelyanskiy, Quantum Information Processing \textbf{14}(1), 1 (2015)

\bibitem{Venturelli15}
D.~Venturelli, S.~Mandr{\`a}, S.~Knysh, B.~O’Gorman, R.~Biswas,
  V.~Smelyanskiy, Phys. Rev. X \textbf{5}(3), 031040 (2015).
\newblock \doi{10.1103/PhysRevX.5.031040}

\bibitem{kim2019}
M.~Kim, D.~Venturelli, K.~Jamieson, in \emph{Proceedings of the ACM Special
  Interest Group on Data Communication} (Association for Computing Machinery,
  New York, NY, USA, 2019), SIGCOMM ’19, p. 241–255.
\newblock \doi{10.1145/3341302.3342072}

\bibitem{Marshall19_Pausing}
J.~Marshall, D.~Venturelli, I.~Hen, E.G. Rieffel, Phys. Rev. Applied
  \textbf{11}, 044083 (2019).
\newblock \doi{10.1103/PhysRevApplied.11.044083}

\bibitem{gonzalez2021}
Z.~Gonzalez~Izquierdo, S.~Grabbe, S.~Hadfield, J.~Marshall, Z.~Wang,
  E.~Rieffel, Phys. Rev. Applied \textbf{15}, 044013 (2021).
\newblock \doi{10.1103/PhysRevApplied.15.044013}

\bibitem{Farhi98}
E.~Farhi, J.~Goldstone, S.~Gutmann, M.~Sipser.
\newblock Quantum computation by adiabatic evolution.
\newblock arXiv:quant-ph/0001106 (2000)

\bibitem{Smelyanskiy12}
V.N. Smelyanskiy, E.G. Rieffel, S.I. Knysh, C.P. Williams, M.W. Johnson, M.C.
  Thom, W.G. Macready, K.L. Pudenz.
\newblock A near-term quantum computing approach for hard computational
  problems in space exploration.
\newblock arXiv:1204.2821 (2012)

\bibitem{advantage}
C.~McGeoch, P.~Farr\'e, The d-wave advantage system: an overview.
\newblock Tecnhical report 14-1049A-A, D-Wave Sys (2020)

\bibitem{dwave_ice}
Error sources for problem representation.
\newblock https://docs.dwavesys.com/docs/latest/c\_qpu\_ice.html

\bibitem{Willsch21}
D.~Willsch, M.~Willsch, C.D.G. Calaza, F.~Jin, H.D. Raedt, M.~Svensson,
  K.~Michielsen,   (2021)

\bibitem{finnila1994quantum}
A.~Finnila, M.~Gomez, C.~Sebenik, C.~Stenson, J.~Doll, Chemical physics letters
  \textbf{219}(5-6), 343 (1994)

\bibitem{Choi08}
V.~Choi, Quantum Information Processing \textbf{7}(5), 193 (2008)

\bibitem{Lucas13}
A.~Lucas.
\newblock Ising formulations of many {NP} problems.
\newblock arXiv:1302.5843 (2013)

\bibitem{Shin2014comment}
S.W. Shin, G.~Smith, J.A. Smolin, U.~Vazirani.
\newblock Comment on ``{D}istinguishing classical and quantum models for the
  {D-W}ave device".
\newblock arXiv:1404.6499 (2014)

\bibitem{Pudenz2016}
K.L. {Pudenz}, arXiv e-prints arXiv:1611.07552 (2016)

\bibitem{Johnson2011quantum}
M.~Johnson, M.~Amin, S.~Gildert, T.~Lanting, F.~Hamze, N.~Dickson, R.~Harris,
  A.~Berkley, J.~Johansson, P.~Bunyk, et~al., Nature \textbf{473}(7346), 194
  (2011)

\bibitem{job2018test}
J.~Job, D.~Lidar, Quantum Science and Technology \textbf{3}(3), 030501 (2018)

\bibitem{Boothby2021}
K.~{Boothby}, C.~{Enderud}, T.~{Lanting}, R.~{Molavi}, N.~{Tsai}, M.H.
  {Volkmann}, F.~{Altomare}, M.H. {Amin}, M.~{Babcock}, A.J. {Berkley},
  C.~{Baron Aznar}, M.~{Boschnak}, H.~{Christiani}, S.~{Ejtemaee}, B.~{Evert},
  M.~{Gullen}, M.~{Hager}, R.~{Harris}, E.~{Hoskinson}, J.P. {Hilton},
  K.~{Jooya}, A.~{Huang}, M.W. {Johnson}, A.D. {King}, E.~{Ladizinsky},
  R.~{Li}, A.~{MacDonald}, T.~{Medina Fernandez}, R.~{Neufeld},
  M.~{Norouzpour}, T.~{Oh}, I.~{Ozfidan}, P.~{Paddon}, I.~{Perminov},
  G.~{Poulin-Lamarre}, T.~{Prescott}, J.~{Raymond}, M.~{Reis}, C.~{Rich},
  A.~{Roy}, H.~{Sadeghi Esfahani}, Y.~{Sato}, B.~{Sheldan}, A.~{Smirnov}, L.J.
  {Swenson}, J.~{Whittaker}, J.~{Yao}, A.~{Yarovoy}, P.I. {Bunyk}, arXiv
  e-prints arXiv:2108.02322 (2021)

\bibitem{chien:asp}
S.~Chien, G.~Rabideau, R.~Knight, R.~Sherwood, B.~Engelhardt, D.~Mutz,
  T.~Estlin, B.~Smith, F.~Fisher, T.~Barrett, G.~Stebbins, D.~Tran, in
  \emph{Proceedings of SpaceOps} (2000)

\bibitem{Achlioptas99}
D.~Achlioptas, E.~Friedgut, Random Structures and Algorithms \textbf{14}(1), 63
  (1999)

\bibitem{CojaOghlan13}
A.~Coja-Oghlan.
\newblock Upper-bounding the k-colorability threshold by counting covers.
\newblock arXiv:1305.0177 (2013)

\bibitem{Rieffel2014AAAI}
E.G. Rieffel, D.~Venturelli, I.~Hen, M.~Do, J.~Frank, in \emph{Proceedings of
  the Twenty-Eighth AAAI Conference on Artificial Intelligence (AAAI-14)}
  (2014), pp. 2337 -- 2343

\bibitem{Culberson95hidingour}
J.~Culberson, A.~Beacham, D.~Papp, in \emph{Proceedings of the CP95 Workshop on
  Studying and Solving Really Hard Problems} (1995), pp. 31--42

\bibitem{casestudy}
E.G. Rieffel, D.~Venturelli, B.~O'Gorman, M.B. Do, E.M. Prystay, V.N.
  Smelyanskiy, Quantum Information Processing \textbf{14}(1), 1 (2015)

\bibitem{Cai-14}
J.~Cai, B.~Macready, A.~Roy.
\newblock A practical heuristic for finding graph minors.
\newblock {arXiv:1406:2741} (2014)

\bibitem{Kowalsky21}
M.~Kowalsky, T.~Albash, I.~Hen, D.~Lidar,   (2021)

\bibitem{pegasus_topology}
K.~Boothby, P.~Bunyk, J.~Raymond, A.~Roy, Next-generation topology of d-wave
  quantum processors.
\newblock Technical report 14-1026A-C, D-Wave Sys (2019)

\end{thebibliography}

\clearpage
\onecolumngrid
\appendix

\section{Optimization of $J_F$ over problem sizes}

\begin{figure}[h!]
	\includegraphics[width=0.5\columnwidth]{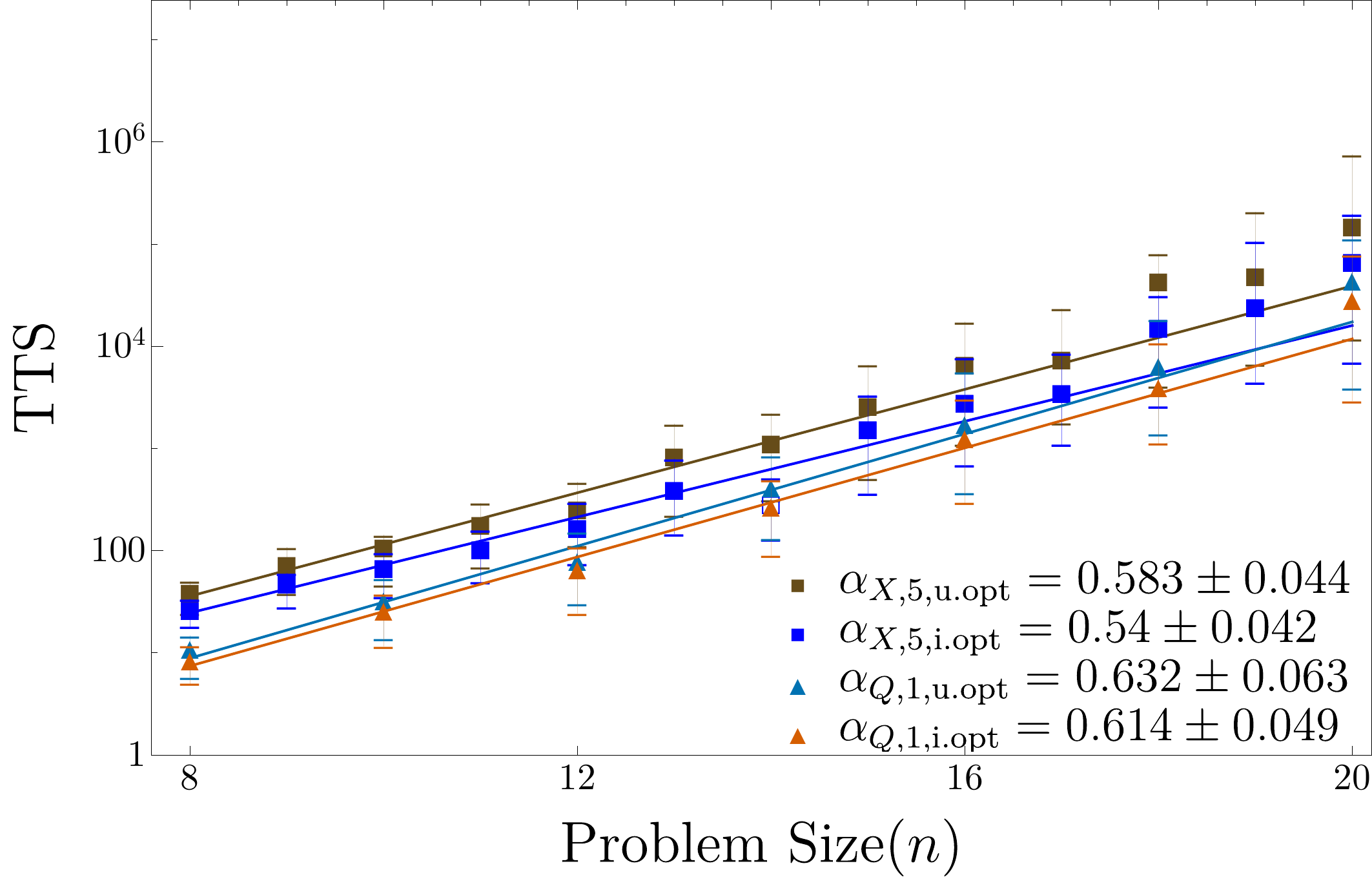}
	\caption{TTS as a function of problem-size for two different $J_F$ optimization methods. `u.opt' and `i.opt' refer to whether $J_F$ was optimized across all the problems of the same size or individually for each problem instance, respectively.}
	\label{fig:ind_vs_uniform_jf}
\end{figure}

\begin{figure}[h]

    \includegraphics[trim=25 0 20 0  , clip, width=\linewidth]{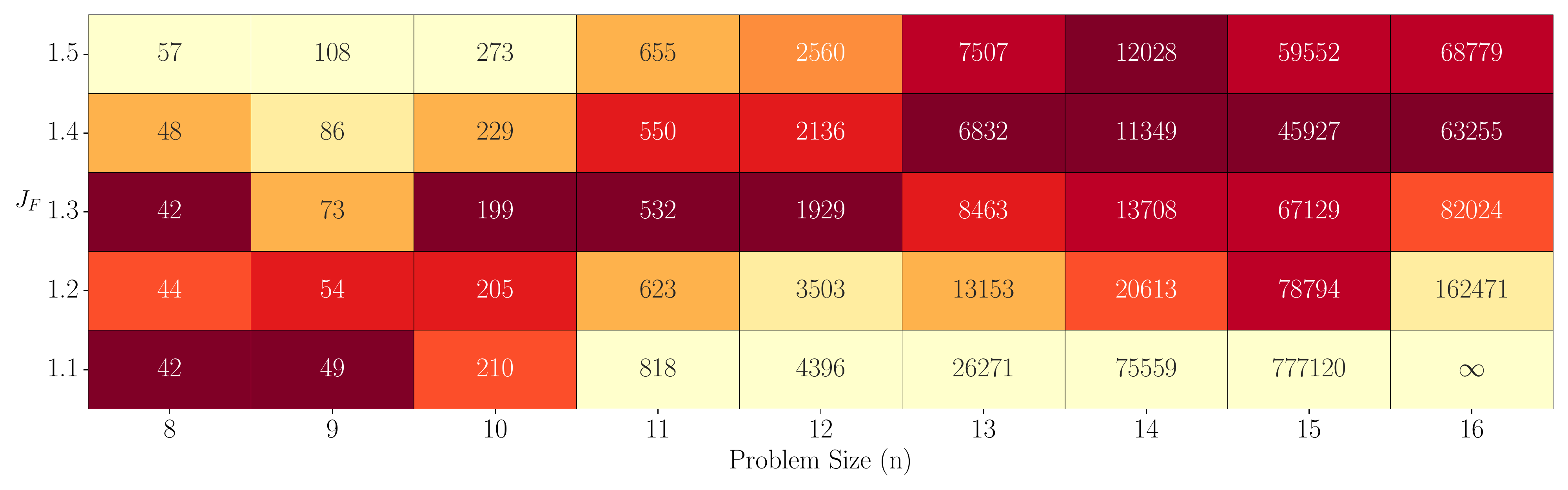}\\
        (a) D-Wave Two \\

	\includegraphics[trim=25 0 20 0  , clip,width=\linewidth]{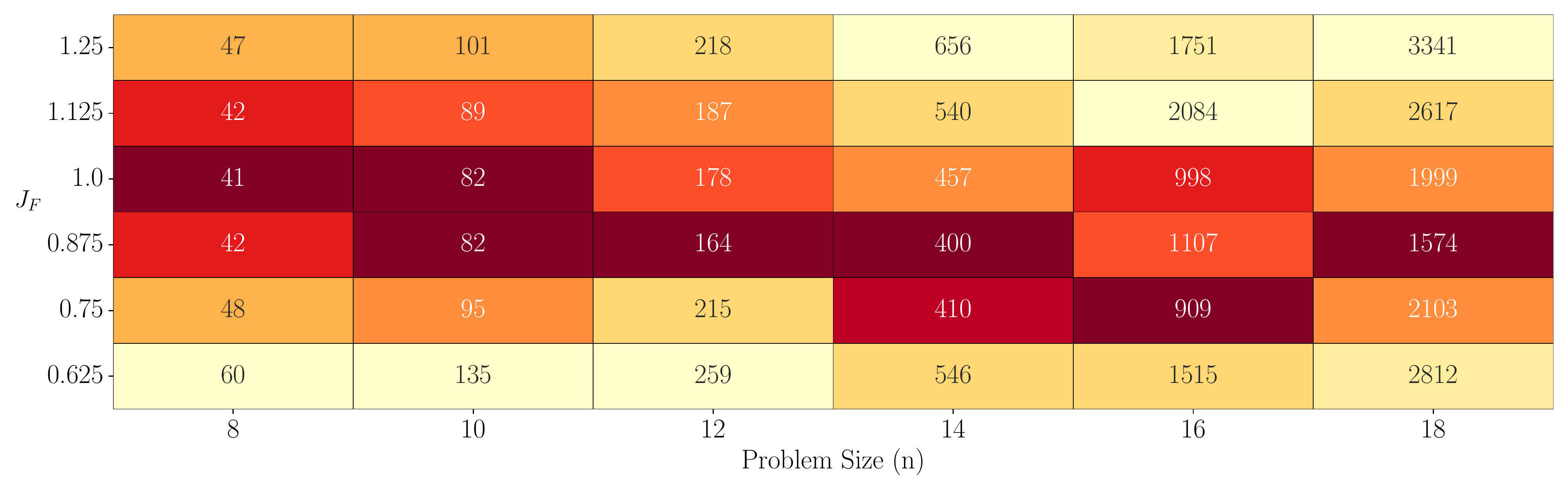}
	\\
	    (b) D-Wave 2X \\

	\includegraphics[trim=25 0 20 0  , clip,width=\linewidth]{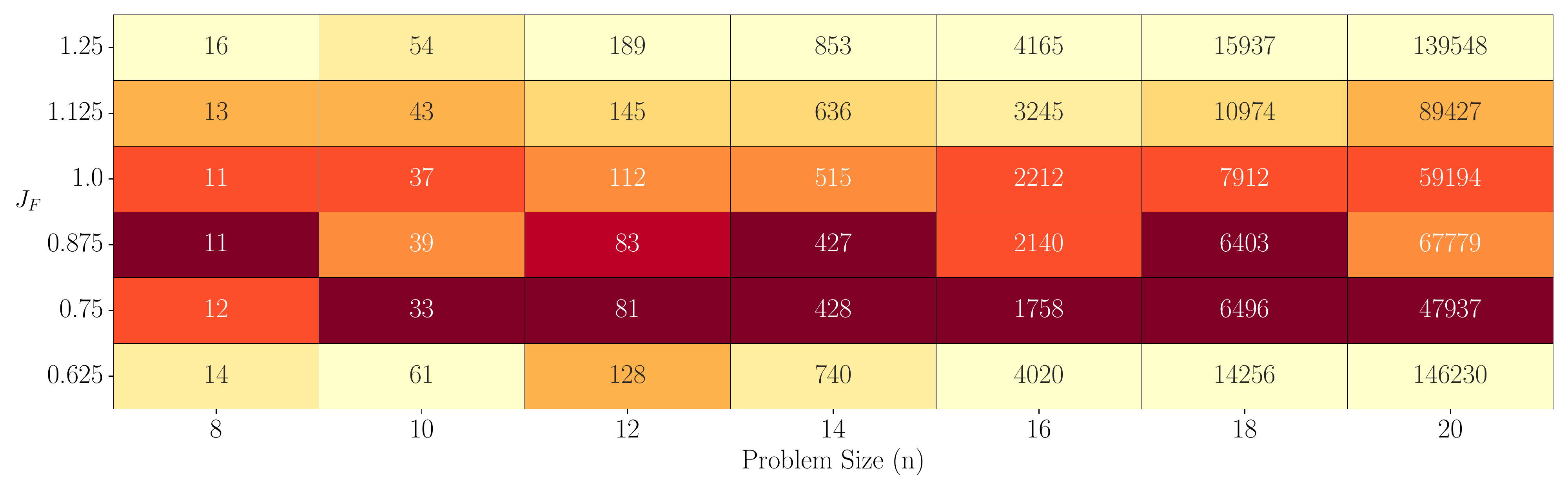} \\
		(c) D-Wave 2000Q\\

	\caption{ Optimal ferromagnetic coupling for different problem sizes. TTS is reported at different coupling parameter values and various graph sizes. The color scaling for each problem size goes from yellow (high $T$) to orange (medium $T$) to red (low $T$).} 

	\label{fig:sizeVsCoupling}
\end{figure}

To optimize the ferromagnetic coupling, we ran each instance at various values of $J_F$. These couplings can be optimized for each vertex model, problem instance, or problem size. Optimizing at the vertex model was ineffective. We tried to use various optimization algorithms for this task. However, we were unable to optimize at the level of vertex models, owing to a large number of parameters and the noisiness of the landscape.  The results reported in this report are for $J_F$ optimization for each instance (`i.opt') unless otherwise specified. Fig. \ref{fig:sizeVsCoupling} shows a heat map of the TTS for each problem size (`u.opt') such that all vertex models of all problem instances have this $J_F$. The reported TTS is averaged over 100 instances. On D-Wave Two, the optimal Jf at each problem size seemed to increase with problem size \cite{casestudy}. This trend does not appear on the 2X and 2000Q. In fact, for these two annealers, the setting $J_F =0.875$ and $0.75$ seems to work well for most problem sizes.  Fig. \ref{fig:ind_vs_uniform_jf} shows how TTS scales under `i.opt' and `u.opt'. On both 2X and 2000Q, we notice a slight reduction in the TTS scaling under `i.opt'.  We do not report $J_F$ optimization for Advantage, which will be studied in detail in other work currently in preparation. Here we did optimize for $J_F \in [-0.4, -1]$ for $n=24$ and $t=20 \mu$s and found $J_F=-0.5$ to be optimal. For the rest of the problem sizes, we only considered $J_F = -1.0$ and $-0.5$; the latter value is close to Advantage's `u.opt' value.

\end{document}